\documentclass[11pt]{article}
\pdfoutput = 1
\usepackage{verbatim}
\usepackage{amssymb,amsfonts,amsthm,amsmath,amssymb,amscd,amstext,epsfig}
\usepackage{color}

\definecolor{darkgreen}{rgb}{0,.5,0}

\def\be{\begin{equation}}
\def\ee{\end{equation}}
\def\sgn{\operatorname{sgn}}
\newcommand{\abs}[1]{\lvert#1\rvert}

\def\Vol{\operatorname{Vol}}
\def\SD{\operatorname{SD}}
\def\tr{\operatorname{tr}}
\def\calN{{\mathcal N}}

\begin{document}

\begin{flushright}
PUPT-2396\\
NSF-KITP-11-220

\end{flushright}

\begin{center}
\vspace{1cm} { \LARGE {\bf Matrix Models for Supersymmetric Chern-Simons Theories with an ADE Classification}}

\vspace{1.1cm}

Daniel R.~Gulotta, J.~P.~Ang,
and Christopher P.~Herzog

\vspace{0.7cm}

{Department of Physics, Princeton University \\
     Princeton, NJ 08544, USA }

\vspace{0.7cm}

{\tt dgulotta, jang, cpherzog@princeton.edu} \\

\vspace{1.5cm}

\end{center}

\begin{abstract}
\noindent
We consider ${\mathcal N}=3$ supersymmetric Chern-Simons (CS) theories that contain product 
$U(N)$ gauge groups and bifundamental matter fields.  Using the matrix model of Kapustin, Willett and Yaakov, we examine the Euclidean partition function of these theories on an $S^3$ in the large $N$ limit.  We show that the only such CS theories for which the long range forces between the eigenvalues cancel have quivers which are in one-to-one correspondence with the simply laced affine Dynkin diagrams.  As the $A_n$ series was studied in detail before, in this paper we compute the partition function for the $D_4$ quiver.  
The $D_4$ example gives further evidence for a conjecture that
the saddle point eigenvalue distribution is determined by the distribution of gauge invariant chiral operators.  We also see that the partition function is invariant under a generalized Seiberg duality for CS theories.

\end{abstract}

\pagebreak

\tableofcontents

\setcounter{page}{1}
\setcounter{equation}{0}

\section{Introduction}

We hope that supersymmetric (SUSY) gauge theories in 2+1 dimensions will help us learn about general features of 2+1 dimensional gauge theories which in turn might shed light on certain condensed matter systems with emergent gauge symmetry at low temperatures.
For many years, there has been strong evidence that, similar to the four dimensional case, three dimensional gauge theories are related to each other by a large number of non-perturbative dualities.  In four dimensions, anomalies provided quantitative tests of these conjectured  dualities, while in three dimensions, because of the absence of anomalies, the situation was more difficult.  Recently, with the introduction of methods for calculating a superconformal 
index on $S^2 \times S^1$
\cite{Kim:2009wb}
 and the Euclidean partition function on $S^3$ 
 \cite{Kapustin:2009kz}, the situation has changed dramatically.     

In this paper, we continue an investigation, 
started in \cite{Herzog:2010hf,Gulotta:2011aa,Gulotta:2011si},
 of the
the large $N$ limit of the $S^3$ partition function 
of supersymmetric (SUSY) Chern-Simons (CS) theories.
The partition function $Z_{S^3}$  is calculated using the matrix model derived in ref.\ \cite{Kapustin:2009kz} by localization (later improved by \cite{Jafferis:2010un,Hama:2010av} to allow matter fields to acquire anomalous dimensions).  
For the CS theory at its superconformal fixed point, the matrix model of \cite{Kapustin:2009kz} computes exactly the partition function and certain supersymmetric Wilson loop expectation values.  
The theories we examine here have ${\mathcal N}=3$ SUSY, a product $U(N)$ gauge group structure, and field content summarized by a quiver diagram.
While in our previous work  \cite{Herzog:2010hf,Gulotta:2011aa,Gulotta:2011si} we examined CS theories that had a $U(N)^d$ gauge group, in this work we relax the constraint that the ranks of the $U(N)$ groups all be equal to each other. 

We examine quantitatively some of these three dimensional dualities.  
We will study the generalized Seiberg duality of  \cite{Aharony:1997gp,Giveon:2008zn}.
Modulo convergence issues, refs.\ \cite{Willett:2011gp,Benini:2011mf} demonstrated that $|Z_{S^3}|$ is invariant under this duality.  In this paper, 
we investigate what this invariance looks like in the large $N$ limit.
We will also examine the AdS/CFT duality between these gauge theories and Freund-Rubin
compactifications of eleven dimensional supergravity.

A big motivation for this paper and our previous work 
\cite{Herzog:2010hf,Gulotta:2011aa,Gulotta:2011si} is the hope that the matrix model can shed 
light on the microscopic origin of the mysterious 
$N^{3/2}$ scaling of the free energy \cite{Klebanov:1996un}
indicated by AdS/CFT.
An early investigation of this matrix model \cite{Drukker:2010nc} demonstrated that the free energy of maximally supersymmetric $SU(N)$ Yang-Mills theory at its infrared fixed point scales as $N^{3/2}$.
By ``free energy'' $F$, we mean minus the logarithm of the partition function on $S^3$, 
\be
F = - \log |Z_{S^3}| \ .
\ee
While one could declare victory, the localization procedure of \cite{Kapustin:2009kz} leaves the microscopic physics of this scaling obscure.  Here we study a larger class of gauge theories using different methods in the hopes of shedding further light on this large $N$ 
scaling.\footnote{%
 See \cite{Marino:2011eh} for a recent Fermi gas interpretation of the microscopic physics.
}

In the context of the AdS/CFT correspondence, the free energy $F$ can be computed from a
classical gravity model dual to the strongly interacting, large $N$ field theory.  For a CFT dual to $AdS_4$ of radius $L$ and effective four-dimensional Newton constant $G_N$, $F$ is given
by \cite{Emparan:1999pm}
\be
F = \frac{\pi L^2}{2 G_N} \ .
\label{emparanresult}
\ee
These types of $AdS_4$ backgrounds arise as Freund-Rubin compactifications of 
$AdS_4 \times Y$ of M-theory, where $Y$ is a seven-dimensional Sasaki-Einstein space supported by $N$ units of four-form flux.  The quantization of $L$ in Planck units implies that at
large $N$ eq.\ (\ref{emparanresult}) becomes \cite{Herzog:2010hf}
\be
F = N^{3/2} \sqrt{\frac{2 \pi^6}{27 \Vol(Y)}} 
 \ .
\label{FVolrel}
\ee
(Here, the volume of $Y$ is computed with an Einstein metric that satisfies the normalization condition $R_{mn} = 6 g_{mn}$.)  
These types of Freund-Rubin solutions arise as the near horizon limit of a stack of $N$ M2-branes placed at the tip of the Calabi-Yau cone $X$ over $Y$.  The CFT in question is then the low energy field theory description of the M2-branes.
The matrix model can be used to compute not only the three-halves scaling
 but also the volume factor and thus allows for quantitative tests of proposed AdS/CFT dual pairs.

Note that the conjecture of ref.\ \cite{Klebanov:1996un} preceded the discovery of AdS/CFT and also concerned the scaling not of $F$ but of the thermal free energy.
AdS/CFT makes clear that the $N^{3/2}$ scaling comes from the over-all normalization of the gravitational action and thus is a universal property of many quantities in the dual field theory.

In trying to elucidate the connection between matrix model quantities and field theory quantities 
in \cite{Gulotta:2011aa,Gulotta:2011si}, we found an interesting relationship between the saddle point eigenvalue distribution and the distribution of gauge invariant operators in the chiral ring.  
We now review this relationship.
The localization procedure of \cite{Kapustin:2009kz} reduces the partition function to an integral over $d$ constant $N \times N$ matrices $\sigma_a$ where $\sigma_a$ is the real scalar that belongs to the same ${\mathcal N}=2$ multiplet as the gauge connection. 
Ref.\ \cite{Herzog:2010hf} 
presented a procedure (later improved in \cite{Jafferis:2011zi}) to evaluate the matrix model by 
saddle point integration in the large $N$ limit.  
At the saddle point, the real parts of the eigenvalues
$\lambda_j^{(a)}$ of $\sigma_a$ grow as a positive power of $N$ while their imaginary parts stay of order one as $N$ is taken to infinity.  Additionally, the real parts of the eigenvalues are the same for each gauge group.  Therefore, in order to find the saddle point, one can consider the large $N$ ansatz
\be
\lambda_j^{(a)} = N^{\alpha} x_j + i y_{a,j} + \ldots
\ee
As one takes $N \to \infty$, the $x_j$ and $y_{a,j}$ become dense; one can introduce an eigenvalue density 
\be
\rho(x) = \lim_{N \to \infty} \frac{1}{N} \sum_{j=1}^N \delta(x-x_j) \ .
\ee

The chiral ring of these quiver gauge theories consists of gauge invariant products of the bifundamental and fundamental matter fields and monopole operators modulo superpotential and monopole relations.  At large $N$, only the so-called ``diagonal monopole operators'' are important.  These operators turn on the same number of units of flux through the diagonal $U(1)$ subgroup of each $U(N)$ gauge group.  Operators in the chiral ring therefore have an associated
R-charge $r$ and a (diagonal) monopole charge $m$.  
To characterize these operators, we introduce the function $\psi(r,m)$ which counts the number of operators in the chiral ring with R-charge less than $r$ and monopole charge less than $m$.  
In \cite{Gulotta:2011aa,Gulotta:2011si}, we found evidence for the following relation between the saddle point eigenvalue distribution and $\psi(r,m)$:
\be
\left. \frac{\partial^3 \psi}{\partial r^2 \partial m} \right|_{m = rx / \mu}
= \frac{r}{\mu} \rho(x) \ , 
\label{evchiralrel}
\ee
where $\mu$ is a Lagrange multiplier that enforces the constraint $\int \rho(x) dx = 1$.  
Defining a function $\psi_{X}(r,m)$ that counts the number of operators without the field $X$, there is a similar conjectured relation  \cite{Gulotta:2011aa,Gulotta:2011si} between $\psi_X^{(1,1)}(r,m)$ and the $y_{a,j}$ which we shall review in section \ref{sec:D4counting}.

\begin{figure}[t]
\begin{center}
 \epsfig{file=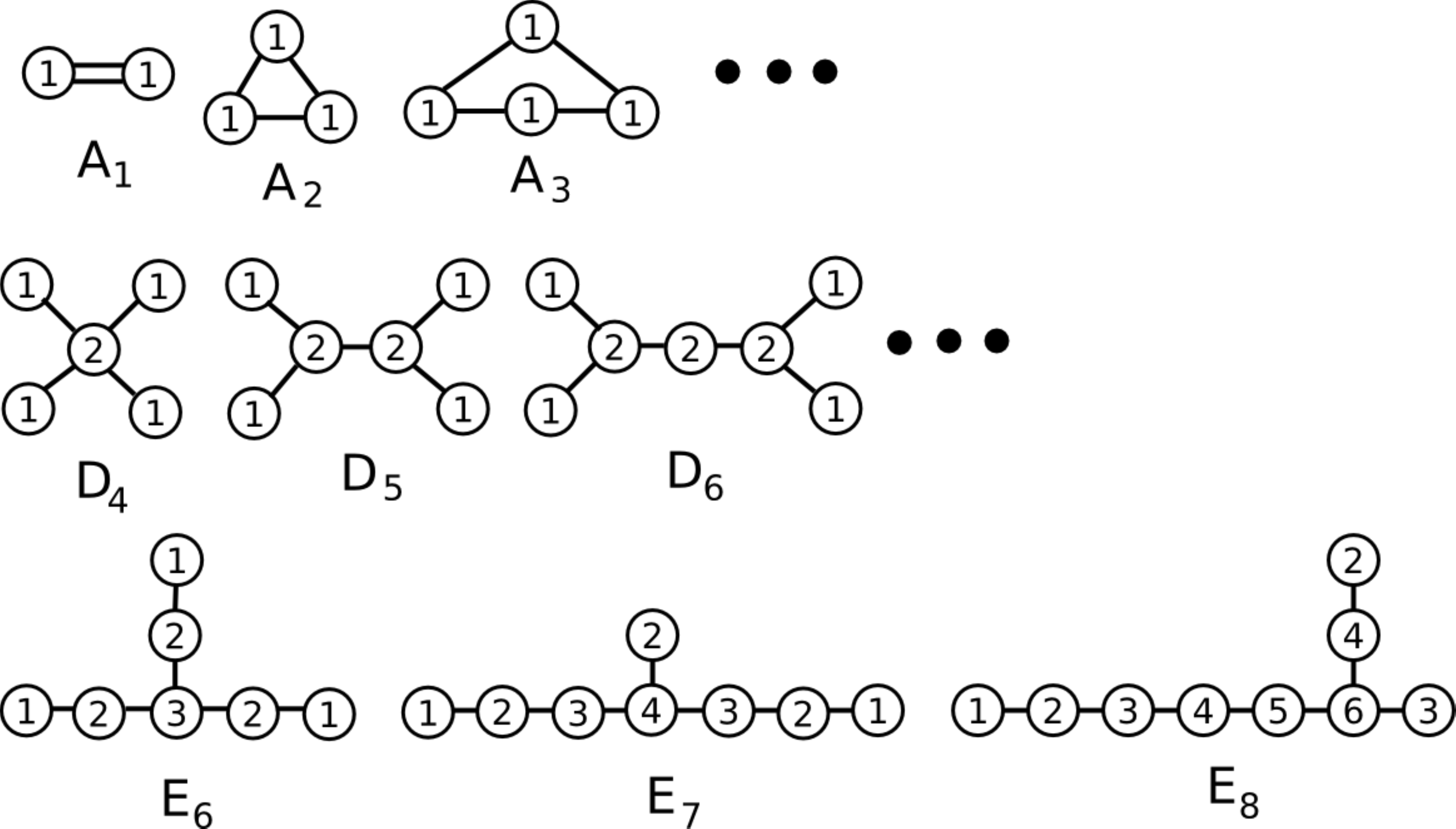,width=5.5in,angle=0,trim=0 0 0 0}%
\end{center}
\caption{
The simply laced affine Dynkin diagrams of ADE type.  \label{fig:dynkin}}
\end{figure}

In the current paper, we begin in section \ref{sec:ADE} by examining CS theories with ${\mathcal N}=3$ SUSY with matter fields only in bifundamental representations of the product gauge group.  However, we relax the constraint that the ranks of the $U(N_a)$ gauge groups all be equal.  We find that the saddle point procedure described in \cite{Herzog:2010hf} for evaluating the matrix model works if 
the ranks of the gauge groups satisfy the following condition:
\be
2 N_a = \sum_{b | (a,b) \in E} N_b \ ,
\label{ADEcond}
\ee
where $E$ is the set of bifundamental fields.  (In other words, the number of flavors for a given gauge group should be twice the number of colors.)
If this condition is not satisfied, there will in general be long range forces between the eigenvalues that render the method in \cite{Herzog:2010hf} inapplicable.
Interestingly, this condition 
implies that the gauge groups and matter content 
are described by a simply laced, affine Dynkin diagram (see figure \ref{fig:dynkin}).

For a $3+1$ dimensional theory, the condition (\ref{ADEcond}) would seem very natural because it implies the vanishing of the NSVZ beta function and the existence of an IR superconformal fixed point.  In this $2+1$ dimensional context, it is less clear where such a condition comes from.
Gaiotto and Witten \cite{Gaiotto:2008ak} call such quivers balanced.   In their study of ${\mathcal N}=4$ Yang-Mills theories in 2+1 dimensions (without CS terms), they find that the Coulomb branch of the moduli space for balanced quivers has an enhanced global symmetry.

When the ranks $N_a$ of the gauge groups are all equal, the Dynkin diagram is of $A_n$ type and was studied in detail in our earlier papers \cite{Herzog:2010hf,Gulotta:2011aa,Gulotta:2011si}.  However, one may also try to solve the matrix model for the $D_n$ and $E_n$ quivers.  
In section \ref{sec:seiberg}, we discuss how generalized Seiberg duality acts on the 
$A_n$, $D_n$, and $E_n$ quivers in the large $N$ limit.  In particular we investigate how the 
saddle point eigenvalue distribution transforms under
Seiberg duality.
In section \ref{sec:D4}, we investigate the $D_4$ case in detail.  
We find that 
the relation (\ref{evchiralrel}) between $\rho(x)$ and the chiral ring continues to hold.
A discussion section contains some partial results and conjectures.
In appendix \ref{app:simple}, we find a simple class of solutions for any ADE quiver.

\vskip 0.1in

\noindent
{\bf Note Added:}
After this paper appeared, we became aware of \cite{Amariti:2011uw}.  Section 4 of \cite{Amariti:2011uw} has some overlap with section \ref{sec:seiberg} of this paper.

\section{${\mathcal N}=3$ SUSY CS Theories and ADE Dynkin diagrams}
\label{sec:ADE}

We are interested in Chern-Simons theories with matter and at least ${\mathcal N}=3$ supersymmetry in 2+1 dimensions.
We assume the gauge group to have the product structure $\prod_{a=1}^d U(N_a)$.
In other words, for each $U(N_a)$ factor, there exists a vector multiplet transforming in the corresponding adjoint representation.  We denote the Chern-Simons level for each $U(N_a)$ factor
$k_a$.
In addition to the vector multiplets, we allow the theory to contain hypermultiplets in bifundamental representations of the gauge groups.  More precisely, in ${\mathcal N}=2$ language, the hypermultiplet consists of a pair of chiral multiplets $(X_{ab}, X_{ba})$ transforming in conjugate bifundamental representations, $(\overline N_a,  N_b)$ and $(\overline N_b,  N_a)$.
The field content of such a theory is conveniently described with a quiver.  For each $U(N_a)$, we draw a vertex.  For each pair $(X_{ab}, X_{ba})$, we join nodes $a$ and $b$ with an edge.  Let $E$ be the set of edges.

Recall that an ${\mathcal N}=3$ vector multiplet contains an ${\mathcal N}=2$ vector multiplet $V$ and a ${\mathcal N}=2$ chiral multiplet $\Phi$ transforming in the adjoint representation of the gauge group.  
The scalar component $\phi$ of $\Phi$ combines with the auxiliary scalar $\sigma$ in $V$ to form a triplet under the $SU(2)_R$ symmetry.  Similarly, the pair $(X_{ab}, X_{ba})$ form a doublet.
The ${\mathcal N}=2$ formalism leaves only a $U(1)_R \subset SU(2)_R$ of the R-symmetry manifest.  Under this $U(1)_R$, $\Phi$ has charge one and $(X_{ab}, X_{ba})$ has charge 1/2.  

We would like to compute the Euclidean partition function of such an ${\mathcal N}=3$ Chern-Simons theory on an $S^3$.  As explained by 
Kapustin, Willett, and Yaakov \cite{Kapustin:2009kz}, this path integral localizes to configurations where the scalars $\sigma_a$ in the vector multiplets are constant Hermitian matrices.  Denoting the eigenvalues of $\sigma_a$ by $\lambda_{a,i}$, $1 \leq i \leq N_a$, the partition function takes the form of the eigenvalue integral
\be
Z = \int \left( \prod_{a, i} d \lambda_{a,i} \right) L_v(\{\lambda_{a,i}\}) L_m(\{\lambda_{a,i} \}) = 
 \int \left( \prod_{a, i} d \lambda_{a,i} \right)  \exp\left[-F (\{ \lambda_{a,i} \}) \right] \ ,
\label{generalZ}
\ee
where the vector multiplets contribute
\be
L_v = \prod_{a=1}^d \frac{1}{N_a!} \left( \prod_{i > j} 2 \sinh[ \pi (\lambda_{a,i} - \lambda_{a,j})] \right)^2
\exp \left(i \pi \sum_{a,j} k_a \lambda_{a,j}^2 \right) \ 
\ee
and the bifundamental matter fields contribute
\be
L_m = \prod_{(a,b) \in E} \prod_{i,j} \frac{1}{2 \cosh [ \pi (\lambda_{a,i} - \lambda_{b,j})]} \ .
\ee

We wish to approximate this integral using the saddle point method in the limit where all of the 
$N_a$ are large.   In particular, we will assume that $N_a = n_a N$ where the $n_a$ are relatively prime integers and $N \gg 1$.
The saddle point equation for the eigenvalue $\lambda_{a,i}$ is
\be
-\frac{1}{\pi} \frac{\partial F}{\partial \lambda_{a,i}} =  2 i k_a \lambda_{a,i} +
\sum_{j \neq i} 2 \coth[\pi (\lambda_{a,i} - \lambda_{a,j})]  -
\sum_{b | (a,b) \in E} \sum_{j} \tanh [ \pi (\lambda_{a,i} - \lambda_{b, j})] =0\ .
\label{saddle point}
\ee

Guided by numerical experiments, we will assume that we can pass to a continuum description where each set of $N_a$ eigenvalues lie along $n_a$ curves, $N$ eigenvalues per curve:
\be
\lambda_{a,I}(x) = N^\alpha x + i y_{a,I}(x) \ ,
\label{evalansatz}
\ee
where $\alpha>0$ and $I = 1, \ldots, n_a$.  The density 
of eigenvalues $\rho(x)$ is assumed to be independent of $a$ and $I$, and we normalize the density such that $\int \rho(x) \, dx = 1$.  
At large $N$, we may use the approximations
\be
\coth[\pi (\lambda_{a,i} - \lambda_{b,j})] \approx \tanh[\pi (\lambda_{a,i} -\lambda_{b,j})] \approx \sgn(x_i -  x_j) \ .
\ee

Assembling these large $N$ approximations, the leading order term in the saddle point equation (\ref{saddle point}) is  (assuming $\alpha < 1$)
\be
\left(2 n_a - \sum_{b | (a,b) \in E} n_b \right) N
\int dx'  \sgn( x-x') \rho(x')   \ .
\ee
For this term to vanish, a sufficient condition is that
\be
2 n_a = \sum_{b | (a,b) \in E} n_b  \ .
\label{confcond}
\ee
To find saddle points for more general theories, one obvious thing to try is a more general ansatz for the eigenvalue distribution, in particular one that does not assume $\rho(x)$ is the same for each gauge group.
 The condition (\ref{confcond}) is surprisingly stringent.  The only allowed quivers that satisfy this condition are the extended ADE Dynkin diagrams!
 In this context, the $n_a$ are called the marks.
 To each node of the Dynkin diagram, we associate a simple root $\alpha_a$.  For an affine Dynkin diagram, we have one additional root $\theta$ associated to the extra node.  This extra root, which is the highest root in the adjoint representation, satisfies the condition $\theta = \sum_a n_a \alpha_a$.  The mark associated with $\theta$ is defined to be equal to one.
 In the rest of the paper we will assume that (\ref{confcond}) holds.

Next we examine the large $N$ limit of $F$ itself, assuming (\ref{confcond}).
It is convenient to separate the free energy into pieces: $F = F_{\rm tree} + F_{\rm loop} + \sum_a \ln (N_a!)$ where
\be
F_{\rm tree} = -i \pi \sum_{a,i} k_a \lambda_{a,i}^2 \ .
\ee
The $\sum_a \ln (N_a!)$ contribution will turn out to be subleading in $N$, and we hereby drop it.
In the large $N$ limit, we find that
\be
F_{\rm tree} =  \pi N  \sum_{a=1}^d \left[ -i N^{ 2\alpha} n_a k_a \int   x^2  \rho(x) dx 
+ 2  N^\alpha \sum_{I=1}^{n_a} k_a  \int  x y_{a,I}(x)\rho(x)dx  +O(1) \right]\ .
\label{Ftree}
\ee

The loop contribution to $F$ is 
\begin{eqnarray}
F_{\rm loop} &\approx &
N^2 \int dx \, dx' \, \rho(x) \rho(x') \times
\nonumber \\
&&
\Biggl(
- \sum_{a=1}^d \sum_{I=1}^{n_a} \sum_{J=1}^{n_a}
 \ln 2 \sinh [ \pi  N^\alpha |x-x'| + i \pi (y_{a,I}(x) - y_{a,J}(x'))\sgn(x-x') ] 
\nonumber
\\
&&
+ \sum_{(a,b) \in E} \sum_{I=1}^{n_a} \sum_{J=1}^{n_b}
\ln 2 \cosh [ \pi N^\alpha (x - x') + i \pi (y_{a,I}(x) - y_{b,J}(x'))] 
\Biggr)  \ .
\end{eqnarray}
This integral is dominated by the region $x \approx x'$ 
which suggests we change variables from $x'$ to $\xi = \pi N^\alpha (x-x')$ so that to a first approximation at large $N$
we may write $y_a(x') = y_a(x- N^{-\alpha} \xi / \pi) \approx y_a(x)$.  The integral reduces to
\begin{eqnarray}
F_{\rm loop} &\approx& \frac{N^{2-\alpha}}{\pi} \int dx \, d\xi \, \rho(x)^2 \times
\nonumber
\\
&&
\Bigl(
- \sum_{a=1}^d \sum_{I=1}^{n_a} \sum_{J=1}^{n_a} \ln 2 \sinh \left[
|\xi| + i \sgn(\xi) \pi(y_{a,I}(x) - y_{a,J}(x)) \right]
\nonumber
\\
&&
 + \sum_{(a,b) \in E} \sum_{I=1}^{n_a} \sum_{J=1}^{n_b}
\ln 2 \cosh \left[\xi + i \pi (y_{a,I}(x)-y_{b,J}(x)) \right] \Bigr) \ .
\end{eqnarray}
Given (\ref{confcond}), the integral over $\xi$ can be performed and is convergent. 
The following two  intermediate results are useful:
\begin{eqnarray*}
\lim_{M \to \infty} 
\left[ \int_{-M}^M \ln 2 \sinh \left[ |x| + \sgn(x) i a \right] \, dx- M^2 \right]
 &=&\frac{\pi^2}{12}  - \frac{1}{4} \arg \left( e^{2 i (a - \pi/2)} \right)^2
\ ,
\\
\lim_{M \to \infty} \left[ \int_{-M}^M  \ln 2 \cosh(x + i a)\, dx - M^2 \right] &=& \frac{\pi^2}{12} 
- \frac{1}{4} \arg \left( e^{2 i a} \right)^2 
\ .
\end{eqnarray*}
The end result is that
\begin{eqnarray}
F_{\rm loop} &=& 
\frac{N^{2- \alpha}}{4 \pi} \int dx \, \rho(x)^2 \Biggl[
\sum_{a=1}^d \sum_{I=1}^{n_a} \sum_{J=1}^{n_a}  \arg \left(e^{2 \pi i (y_{a,I} - y_{a,J} - 1/2)} \right)^2 
\nonumber
\\
&&
\hspace{30mm}
- \sum_{(a,b) \in E} \sum_{I=1}^{n_a} \sum_{J=1}^{n_b} \arg \left( e^{2 \pi i (y_{a,I} - y_{b,J})} \right)^2
\Biggr] \ .
\end{eqnarray}

To find a nontrivial saddle point, we have two choices.  We can balance $N^{2-\alpha}$ against the first term in (\ref{Ftree}), in which case $\alpha = 1/3$.  Alternately, we can assume that $\sum_a n_a k_a = 0$ and balance $N^{2-\alpha}$ against the second term of (\ref{Ftree}) in which case $\alpha = 1/2$. The first case has a massive type IIA supergravity dual \cite{Jafferis:2011zi}.\footnote{%
 The ansatz (\ref{evalansatz}) needs to be slightly modified in this case to allow both the real and imaginary parts of $\lambda$ to scale with $N^{1/3}$.  See \cite{Jafferis:2011zi}.
}
  We shall make the second choice, leading to an eleven dimensional supergravity dual.

The free energy that we must maximize is then
\begin{eqnarray}
\label{finalF}
F &=& N^{3/2}   \int \rho(x) \Biggl[ 2 \pi x \sum_a \sum_{I=1}^{n_a} k_a y_{a,I}(x)
+ \frac{ \rho(x)}{4\pi} \Biggl(
\sum_{a=1}^d \sum_{I=1}^{n_a} \sum_{J=1}^{n_a}  
\arg \left(e^{2 \pi i (y_{a,I} - y_{a,J} - 1/2)} \right)^2
\nonumber
\\
&&
- \sum_{(a,b) \in E} \sum_{I=1}^{n_a} \sum_{J=1}^{n_b} 
 \arg \left( e^{2 \pi i (y_{a,I} - y_{b,J})} \right)^2 
 \Biggr)
\Biggr] dx - 2 \pi \mu N^{3/2} \left( \int  \rho(x) \, dx- 1 \right) 
\end{eqnarray}
where we have added a Lagrange multipler $\mu$ to enforce the constraint that the eigenvalue density integrates to one.

We believe that the free energy can always be written in a way such that it depends only on the average values 
\be
\overline y_a = \frac{1}{n_a} \sum_I y_{a,I}
\ee
 rather than on the $y_{a,I}$ individually.  In particular, we conjecture that $F$ will take the form
 \begin{eqnarray}
 F &=& \pi N^{3/2} \int \rho(x) 
 \left[ 2  x \sum_a k_a n_a \overline y_a   
 + \rho(x) \sum_{(a,b)\in E} n_a n_b \left( \frac{1}{4} - (\overline y_a - \overline y_b)^2\right) \right] dx \nonumber \\
 &&
 - 2 \pi \mu N^{3/2} \left( \int \rho(x) \, dx -1 \right) \ .
 \label{conjF}
 \end{eqnarray}
 We have verified this conjecture in the case of the $A_n$ and $D_4$ quivers.
 (Indeed, in the $A_n$ case, this result was given in \cite{Gulotta:2011aa,Gulotta:2011si}.)
Note we are not claiming that the differences $y_{a,I} - y_{b,J}$ remain bounded between $\pm 1/2$.  Indeed, for the $D_4$ case, the differences are not bounded.  We are arguing that a more subtle cancellation happens among the integer shifts introduced in evaluating the arg functions such that
(\ref{finalF}) reduces to (\ref{conjF}).\footnote{%
 The second term of (\ref{finalF}) and the second term of (\ref{conjF}) do not have the same periodicity properties as functions of the $y_{a,I}$.
In the limit $\overline y_a - \overline y_b \to \infty$, we do not expect (\ref{finalF}) and (\ref{conjF}) to be equal.  What we are claiming is a region in $y_{a,I}$ space where 
(\ref{finalF}) and (\ref{conjF}) are equal and which contains the saddle point eigenvalue distribution.   
}

\section{Generalized Seiberg Duality}
\label{sec:seiberg}

Modulo some subtle convergence issues,  
refs.\ \cite{Willett:2011gp,Benini:2011mf} demonstrated that the absolute value $|Z_{S^3}|$ of the partition function (\ref{generalZ}) is invariant under a generalized Seiberg duality 
\cite{Aharony:1997gp,Giveon:2008zn}. 
For a $U(N_c)$ theory with $N_f$ flavors and CS level $k$, this generalized Seiberg duality acts by sending $N_c \to N_f - N_c + |k|$ and $k \to -k$. 
In the context of our quiver theories, Seiberg duality can be performed on any individual node of the quiver.  The bifundamental fields look like flavors from the point of view of the node to be dualized.    
We have the long range force cancellation condition (\ref{confcond}) which implies that $N_f = 2 N_c$.  We also work in the large $N$ limit which implies that $k \ll N_f, N_c$.  Thus, under Seiberg duality $N_c$ is invariant to leading order in $N$.

In the quiver context, in addition to changing the CS level of the dualized node, the duality also changes the CS levels of the neighboring nodes.  If we dualize node $a$, 
then for $b | (a,b) \in E$, $k_b \to k_b + k_a$. 
Note that the sum $\sum_a n_a k_a$ is invariant under the duality given (\ref{confcond}). 
To motivate why the CS levels change in the way they do, we recall the brane construction of the $A_n$ quivers  \cite{Kitao:1998mf,Bergman:1999na}.  
We start with type IIB string theory with the 6 direction periodically identified.
We place $N$ D3-branes in the 0126 directions.  Intersecting these D3-branes at intervals in the 6 direction, we insert $(1,p_a)$ 5-branes that span the 012 directions as well as lines in the 37, 48, and 59 planes that make angles $\theta_a = \arg(1 + ip_a)$  
with the 3, 4, and 5, axes respectively.  The choice of $\theta_a$ guarantees that the construction has the right number of supercharges for the 2+1 dimensional gauge theory living on the D3-branes to have ${\mathcal N}=3$ supersymmetry.  
Each unbroken D3-brane interval in the 6 direction corresponds to a $U(N)$ gauge group in the $A_n$ quiver.  The CS level of the gauge group is given by the difference $p_{a}-p_{a+1}$.  
Seiberg duality corresponds to interchanging neighboring $(1,p)$ 5-branes.\footnote{%
 A similar brane construction involving an orbifold should exist for the $D_n$ quivers
 \cite{Gaiotto:2008ak, hep-th/9806238}.
} 

For the simply laced affine Dynkin diagrams $A_n$, $D_n$, and $E_n$, Seiberg duality can be reinterpreted as the action of the Weyl group.  
(There is a similar story for 3+1 dimensional gauge theories with ${\mathcal N}=1$ SUSY \cite{Cachazo:2001sg}.)
To $n$ of the  gauge groups we associate the corresponding simple root $\alpha_a \in {\mathbb R}^n$ of the Dynkin diagram. The $n+1$st gauge group is assigned the root $\alpha_{n+1} = -\theta$ where 
$\theta$ is the highest root.  
The CS level is $k_a = \alpha_a \cdot p$ where $p$ is an arbitrary weight vector.
Note that
\be
\sum_a n_a k_a = - \theta \cdot p+ \sum_{a=1}^n n_a \, \alpha_a \cdot p  = 0 \ ,
\ee
since $\sum_{a=1}^n n_a \alpha_a = \theta$.  (The mark associated with $\theta$ is defined to be one.)

Returning to the brane construction for the $A_n$ quivers, the Weyl group is the permutation group acting on the $(1,p_a)$ branes.  We start by assigning to the $n+1$ gauge groups 
the $n$ simple roots $\alpha_a = e_a  - e_{a+1}$ with $e_a$ a unit coordinate vector in ${\mathbb R}^{n+1}$ and the highest root $-\theta = e_{n+1} - e_1$.  Note that these $n+1$ 
roots live in a ${\mathbb R}^n$ subspace of ${\mathbb R}^{n+1}$.  Let 
$p = (p_1, p_2, \ldots, p_{n+1})$ be a weight vector corresponding to the charges of the 5-branes.  
As claimed, the CS levels are then $k_a = \alpha_a \cdot p$.

The results of refs.\ \cite{Willett:2011gp,Benini:2011mf} imply that $|Z_{S^3}|$ is invariant under this generalization of Seiberg duality for any $N$.  The large $N$ limit (\ref{finalF}) must also be invariant, but 
it is still interesting to see how the invariance arises in our case.  If we shift $k_a \to -k_a$, $k_b \to k_b + k_a$  for $b | (a,b) \in E$, then the first
term of \eqref{finalF} can be made invariant if we take
\be
\sum_I y_{a,I} \to -\sum_I y_{a,I} + \sum_{b|(a,b) \in E} \sum_{J=1}^{n_b} y_{b,J} \ .
\label{impreciseSD}
\ee
In the case when $n_a = 1$, it is clear that not just the first term of (\ref{finalF}) but all of (\ref{finalF}) is invariant under such a transformation.  For example, for the $A_n$ quivers, the transformation on $y_{a,I}$ reduces to
\be
y_a \to - y_a + y_{a-1} + y_{a+1} \ .
\ee
To understand Seiberg duality on the internal nodes of the $D_n$ and $E_n$ quivers, we should
describe how each $y_{a,I}$ transforms rather than how the average transforms. 
However, it is straightforward to demonstrate that the conjectured form of the free energy
(\ref{conjF}) is invariant under (\ref{impreciseSD}).  Thus, 
 if we believe (\ref{conjF}), i.e.\ that the free energy depends only on the average values
$\overline y_a$, then $F$ is invariant under our generalized Seiberg duality.

We should note that our large-$N$ approximation is insensitive to shifting
all of the $y$'s by the same amount.  The true Seiberg duality is actually
a combination of \eqref{impreciseSD} with such a shift.
(See section \ref{sec:numerics} for an example of this phenomenon.)

\section{$D_4$ type quiver}
\label{sec:D4}

Let us examine the $D_4$ type quiver.  
Let the external vertices have CS levels $k_a$, $a=1, \ldots, 4$ and the internal vertex have CS level $k$ where $2k = -\sum_a k_a$.  We label the imaginary part of the eigenvalues of the internal vertex $y_{0,1}(x)$ and $y_{0,2}(x)$ and define $\delta y_{a,I}(x) = y_a(x) - y_{0,I}(x)$.  
Based on numerical simulation and intuition from the $A$-type quivers \cite{Herzog:2010hf}, we expect the saddle point eigenvalue distributions are piecewise smooth functions of $x$.

The action of generalized Seiberg duality separates the space of CS levels into different Weyl chambers.  It is enough to analyze the theory in a given Weyl chamber.  Instead we will make the choice $k_1 \geq k_2 \geq k_3 \geq k_4 \geq 0$ which includes three different Weyl chambers.  At the end, we will be able to see explicitly how Seiberg duality acts on the solution.

In the central region $\abs{x} \leq \mu / 2 k_1$, the equations of motion yield
\be
\rho(x) = \frac{\mu}{2} \ ; \; \; \;
\delta y_{a,I} (x) = \frac{k_a x }{2 \rho(x)} \ .
\ee
Once $\abs{x} > \mu / 2 \abs{k_1}$, the solution gets more complicated.  
In general one of the $\delta y_{1,I}$, which for convenience we choose to be $\delta y_{1,1}$, will saturate at $1/2$ and the other will 
continue to grow such that $\delta y_{1,2} > 1/2$.
There are three sub-cases to consider: 1) $k_1 -k_4 \geq k_2 + k_3$, 2) $k_1 + k_4 \geq k_2 + k_3 \geq k_1 - k_4$ and 3) $k_2 + k_3\geq k_1 + k_4$.  
Let $\delta y_a \equiv y_a - (y_{0,1} + y_{0,2})/2$.    
The next three piecewise linear regions of the solution are
\begin{eqnarray}
\frac{\mu}{2 k_1} \leq x \leq \frac{\mu}{k_1+k_2} &:&
\rho = \frac{\mu}{2} \ , \\
&&
\delta y_{1,1}= \frac{1}{2} \ , 
 \; \; \;
\delta y_2 = \frac{k_2 x}{2 \rho} \ ,  \; \; \; 
\delta y_3 = \frac{k_3 x}{2 \rho} \ ,  \; \; \; 
\delta y_4 = \frac{k_4 x}{2 \rho} 
 \ ,
\nonumber \\
&&
y_{0,1} - y_{0,2} = \frac{k_1 x}{\rho}-1 \ ,
\nonumber \\
\frac{\mu}{k_1+k_2} \leq x \leq \frac{\mu}{k_1+k_3} &:&
\rho = \mu- \frac{(k_1+k_2)x}{2}\ , \\
&&
\delta y_{1,1}= \frac{1}{2} \ , \; \; \;
\delta y_{2,2} = \frac{1}{2}  \ , \; \; \;
\delta y_3 = \frac{k_3 x}{2 \rho} \ ,  \; \; \; 
\delta y_4 = \frac{k_4 x}{2 \rho} \ , 
\nonumber \\
&&
y_{0,1} - y_{0,2} = \frac{(k_1-k_2) x}{2\rho} \ ,
\nonumber \\
\frac{\mu}{k_1+k_3} \leq x \leq x_4 &:&
\rho = \frac{3\mu}{2} - \frac{(2k_1+k_2+k_3)x}{2}\ , \\
&&
\delta y_{1,1}= \frac{1}{2} \ , \; \; \;
\delta y_{2,2} = \frac{1}{2}  \ , \; \; \;
\delta y_{3,2} = \frac{1}{2} \ ,  \; \; \; 
\delta y_4 = \frac{k_4 x}{2 \rho} \ , 
\nonumber \\
&&
y_{0,1} - y_{0,2} = \frac{1}{3} + \frac{(k_1-k_2-k_3) x}{3\rho} \ ,
\nonumber 
\end{eqnarray}
where in cases 1 and 2, $x_4 = \mu /  (k_1+k_4)$ and in case 3, $x_4 = \mu/  (k_2+k_3)$.
The fifth and final piecewise linear region of the solution is different in each of the three cases.  In cases 1 and 2, we find
\begin{eqnarray}
\frac{\mu}{k_1+k_4} \leq x \leq x_5 &:&
\rho = 2\mu- \frac{(3k_1+k_2+k_3+k_4)x}{2}\ , \\
&&
\delta y_{1,1}= \frac{1}{2} \ , \; \; \;
\delta y_{2,2} = \frac{1}{2}  \ , \; \; \;
\delta y_{3,2} = \frac{1}{2} \ ,  \; \; \; 
\delta y_{4,2} = \frac{1}{2} \ , 
\nonumber \\
&&
y_{0,1} - y_{0,2} = \frac{1}{2} - \frac{(k + k_2 + k_3 + k_4) x}{2\rho} \ ,
\nonumber 
\end{eqnarray}
where $x_5 = \mu /  k_1$ in case 1 and $y_{0,1} - y_{0,2} \to 1$ at the endpoint.
In constrast, in case 2, $x_5 = 2 \mu / (k_1 +k_2 + k_3 + k_4)$ and $y_{0,1} - y_{0,2} \to 0$ at the endpoint.
In case 3, we have instead
\begin{eqnarray}
\frac{\mu}{k_2+k_3} \leq x \leq \frac{2\mu}{k_1+k_2+k_3+k_4} &:&
\rho = 2\mu- (k_1+k_2+k_3)x\ , \\
&&
\delta y_{1,1}= \frac{1}{2} \ , \; \; \;
\delta y_{2,2} = \frac{1}{2}  \ , \; \; \;
\delta y_{3,2} = \frac{1}{2} \ ,  \; \; \; 
\delta y_{4,1} = \frac{k_4 x}{2 \rho} \ , 
\nonumber \\
&&
y_{0,1} - y_{0,2} = 0 \ .
\nonumber 
\end{eqnarray}
In the three cases, the value of the Lagrange multiplier is
\begin{eqnarray}
\mbox{case 1} & : &
\frac{2}{\mu^2} = \frac{6}{k_1} + \frac{2 k}{k_1^2} - \frac{1}{k_1+k_2} - \frac{1}{k_1+k_3} - \frac{1}{k_1+k_4} \ , \\ 
\mbox{case 2} & : &
\frac{2}{\mu^2} = -\frac{6}{k} - \frac{2 k_1}{k^2} - \frac{1}{k_1+k_2} - \frac{1}{k_1+k_3} - \frac{1}{k_1+k_4}  \ , \\
\mbox{case 3} & : & 
\frac{2}{\mu^2} = -\frac{4}{k} + \frac{2 k_4}{k^2} - \frac{1}{k_1+k_2} - \frac{1}{k_1+k_3} - \frac{1}{k_2+k_3} \ .
\end{eqnarray}

A virial theorem proven in \cite{Gulotta:2011si} asserts that 
$F = 4 \pi N^{3/2} \mu/3$, and hence from (\ref{FVolrel}) it follows that $\operatorname{Vol}(Y) = \pi^4 / 24 \mu^2$. 
We can be more precise about what exactly the seven dimensional manifold 
$Y$ is.  It is a tri-Sasaki Einstein space.  
A tri-Sasaki Einstein space can be defined as the level surface of a conical manifold with
hyperk\"ahler structure \cite{Boyer:1998sf}.   The eight dimensional hyperk\"ahler cone in turn can be thought of as a hyperk\"ahler quotient \cite{Hitchin:1986ea} 
of the the quaternionic manifold ${\mathbb H}^8$ by a subgroup
$SU(2) \times U(1)^3 \subset U(2) \times U(1)^4$.  
We can define the two special $U(1)$'s that do not appear in the quotient using the gauge connections ${\mathcal A}_a$ and ${\mathcal A}$ of the $U(1)^4$ and $U(2)$ groups respectively. 
One is the diagonal subgroup $\tr {\mathcal A} + \sum_a {\mathcal A}_a$.  The other is
picked out by the CS levels, $k \tr {\mathcal A} + \sum_a k_a {\mathcal A}_a$.   
See for example \cite{Martelli:2008si,Jafferis:2008qz} for more details.

These three cases are related by the Seiberg duality described in the previous section.
Consider a Seiberg duality on the central node, then on node 1, and then again on the central node, $\SD_1 \SD_c \SD_1$.  Under such a transformation the CS levels change such that
\be
(k_1, k_2, k_3, k_4)  \to 
\frac{1}{2} (k_1 + k_2 + k_3 + k_4, k_1 + k_2 -k_3 - k_4, k_1 - k_2 +k_3 - k_4, k_1 -k_2 -k_3 +k_4) \ .
\ee 
Given these relations, it is straightforward to verify that case one is mapped to case two and vice versa.  The map $\SD_4 \SD_1 \SD_c \SD_1$ takes case three into case one with the rule
\be
(k_1, k_2, k_3, k_4)  \to 
\frac{1}{2} (k_1 + k_2 + k_3 + k_4, k_1 + k_2 -k_3 - k_4, k_1 - k_2 +k_3 - k_4, k_1 -k_2 -k_3 +k_4) \ .
\ee

To express the action of Seiberg duality in terms of the Weyl group, we need to introduce simple roots for $D_4$: $\alpha_1 = (1,1,0,0)$, $\alpha_2 = (-1,1,0,0)$, $\alpha_3 = (0,-1,1,0)$, and $\alpha_4 = (0,0,-1,1)$.  Note that $\alpha_3$ corresponds to the central node.  The highest root is then $\theta = (0,0,1,1)$ and the CS levels can be expressed in terms of auxiliary variables $p_a$ such that
\be
k_1 = p_1 + p_2 \ , \; \; \;
k_2 = -p_1 + p_2 \ , \; \; \;
k_3 = -p_3 + p_4 \ , \; \; \;
k_4 = -p_3 - p_4 \ .
\ee
The Weyl group acts by permuting the $p_a$ and by changing the sign of an even number 
of the $p_a$.\footnote{%
$D_4$ also has an outer automorphism group isomorphic
to $S_3$, corresponding to permutations of three of the $k$'s.
These automorphisms cannot be expressed in terms of Seiberg dualities.
Swapping $k_1$ and $k_2$ negates $p_1$; other permutations
act less nicely on the $p_i$.
 Naively, there is an $S_4$, but the four dimensional subgroup generated by the swaps $(12)(34)$, $(13)(24)$, and $(14)(23)$ are Seiberg dualities.
 }
We can write a more general expression for $\mu$, assuming $\abs{p_1} \geq \abs{p_2} \geq \abs{p_3} \geq \abs{p_4}$:
\begin{eqnarray}
\frac{2}{\mu^2} &=& 
- \frac{1}{2 \abs{p_1}} + \frac{6}{\abs{p_1} + \abs{p_2}} - 2\frac{ \abs{p_1} + \abs{p_3}}{(\abs{p_1} + \abs{p_2})^2} +
\\
\nonumber &&
- 
\frac{1}{\abs{p_1} + \abs{p_2} + \abs{p_3} + \abs{p_4}} - \frac{1}{\abs{p_1} + \abs{p_2} + \abs{p_3} - \abs{p_4}} \ .
\end{eqnarray}

\subsection{Comparison with the Numerical Saddle Point}
\label{sec:numerics}

\begin{figure}[t]
\begin{center}
 \epsfig{file=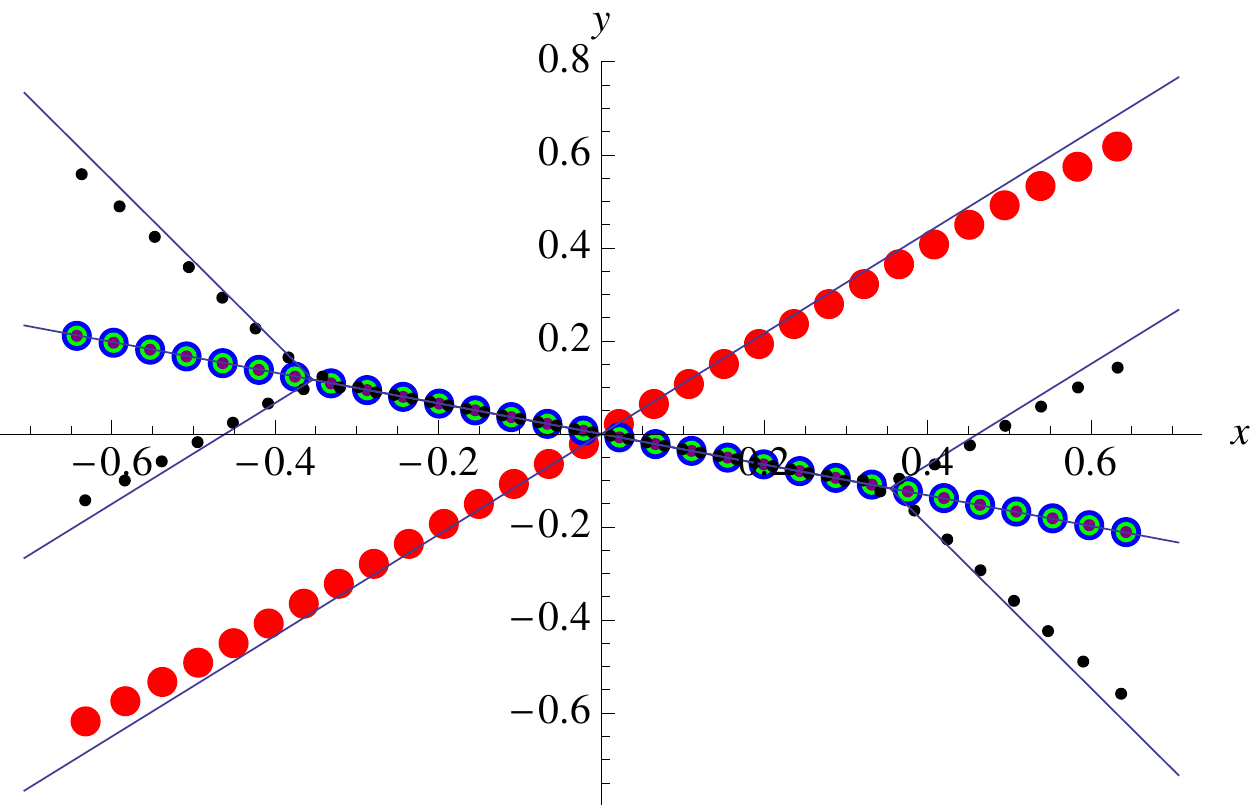,width=2.7in,angle=0,trim=0 0 0 0}%
\hspace{0.2cm}
 \epsfig{file=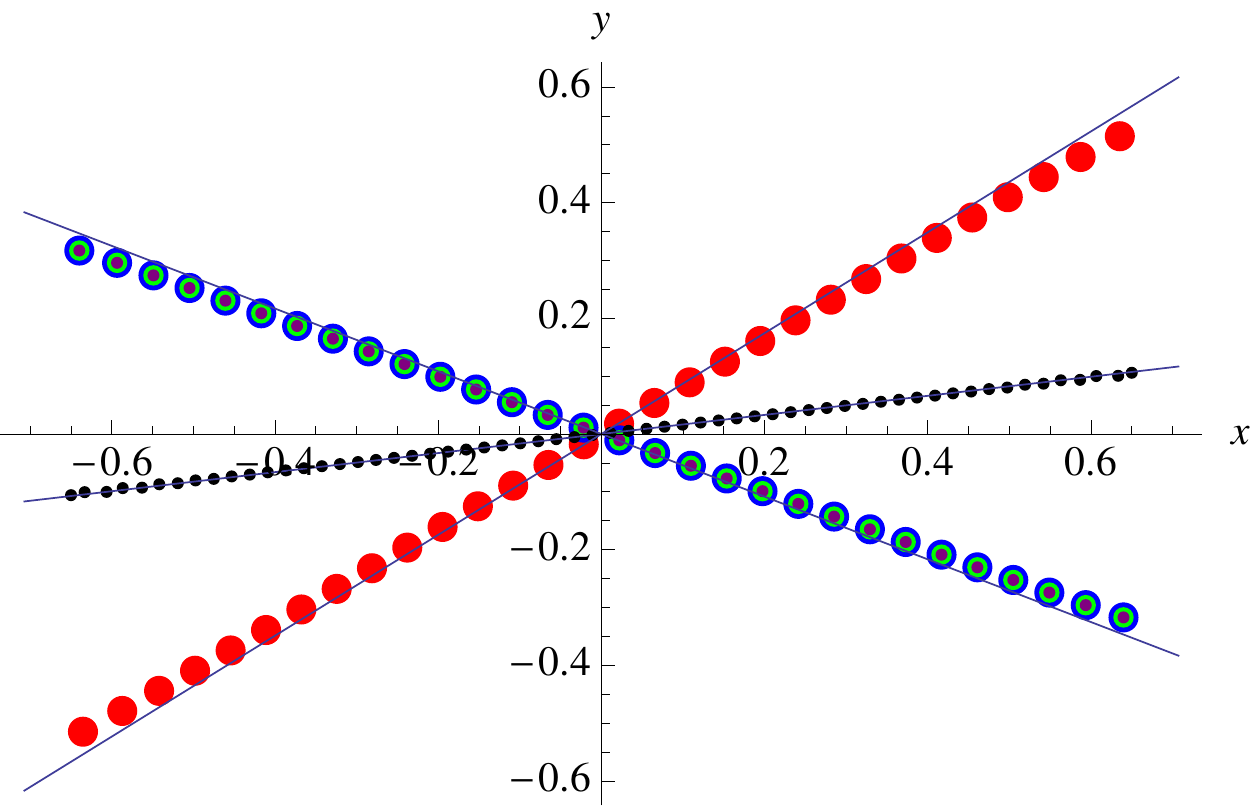,width=2.7in,angle=0,trim=0 0 0 0}%
\end{center}
\caption{
The eigenvalue distribution for two Seiberg dual $D_4$ theories with $N=30$ and $(k_1, k_2, k_3, k_4)=(2,0,0,0)$ (left) or $(1,-1,-1,-1)$ (right).  
The small black points correspond to the eigenvalues of the $U(2N)$ node.
The large red points correspond to the $k_1>0$ node.  The eigenvalues for the remaining gauge groups -- indicated by the blue, green and purple points -- are coincident.
The thin black lines are the large $N$ analytic prediction with the additional input $\overline y_0 \approx -x/3$ (left) and $\overline y_0 \approx x/6$ (right).
  \label{fig:comptonum}}
\end{figure}

To check our large $N$ approximation, we can use a computer to calculate the eigenvalue distributions numerically.  
We use the same relaxation algorithm described in \cite{Herzog:2010hf}.  
For the simple Seiberg dual cases $(k_1, k_2, k_3, k_4) = (2,0,0,0)$ and $(1,-1,-1,-1)$, we plot the answer in figure \ref{fig:comptonum}.  Note that these cases are somewhat degenerate.  Instead of having five distinct piecewise linear regions (for $x>0$), we have only two for the $(2,0,0,0)$ theory and one for the $(1,-1,-1,-1)$ theory.  For $(2,0,0,0)$, 
the last three regions collapse to the point $x =  \mu / k_1$, while for $(1,-1,-1,-1)$ the last four regions collapse to the point $x = \mu / 2 k_1$.  

The thin black lines are the analytic, large $N$ prediction.  The agreement is reasonably good and should improve as $N$ is increased.  Note however that we predict only the differences between the $y_a$.  To make the plot, we first fit $\overline y_0 = (y_{0,1} + y_{0,2})/2$ to a straight line.
Given the value of the fit ($\overline y_0 \approx -x/3$ for $(2,0,0,0)$ and $\overline y_0 \approx x/6$ for $(1,-1,-1,-1)$),
 the transformation (\ref{impreciseSD}) is obeyed only up to an overall additive shift of all the $y_a$.

\subsection{Operator counting}
\label{sec:D4counting}

We would like to count the number of gauge invariant chiral operators of the $D_4$ quiver when $N=1$. 
To that end, we should first construct words out of the bifundamental fields $A_i^\mu$ and $B_{j \nu}$ that are invariant under the $SU(2) \subset U(2)$ gauge group.  The indices $i,j = 1, \ldots, 4$ correspond to the four $U(1)$ gauge groups while $\mu, \nu = 1,2$ index the $U(2)$ group. Then we quotient by the superpotential (or moment map) relations:
\begin{eqnarray}
\label{firstABrel}
\sum_i A_i^{\mu} B_{i \nu} + k \phi^{\mu}_{\nu} & = & 0 \ , \\
\label{secABrel}
A_i^{\mu} B_{i \mu} + k_i \phi_i & = & 0 \ , \\
\phi_i \delta^{\mu}_{\nu} + \phi^{\mu}_{\nu} & = & 0 \ .
\end{eqnarray}
We use the convention that we sum over repeated $U(2)$ indices,
but we do not sum over repeated $U(1)$ indices.
Finally, we restrict to operators with $U(1)$ charges that can be canceled by the corresponding diagonal monopole operators which have charge $(m k_1, m k_2, m k_3, m k_4)$ under the $U(1)^4$, where $m$ is an integer.   (There is a sense in which what we do here is a generalization of the analysis performed in 
\cite{Albertsson:2000px}.)

Our counting problem is equivalent to a much simpler counting problem where we 
ignore the  $\phi_i$ and $\phi^\mu_\nu$ fields.  
Let $R$ denote the ring of operators.  For any $R$-module $M$, let $h(M)$
denote the Hilbert series of $M$.  For any ideal $I$, we have
\begin{equation}
h(R) = h(I) + h(R/I).
\end{equation}
If we let $I = (\phi)$, then $R$ and $I$ are isomorphic $R$-modules and
$h((\phi)) = t h(R)$.  So
\begin{equation}
h(R) = (1-t)^{-1} h(R/(\phi)).
\end{equation}
In terms of the function $\psi(r,m)$ defined in the introduction, now with a subscript to indicate the choice of ring, we have from (\ref{evchiralrel})
\begin{equation}
\rho(x) = \frac{\mu}{r} \psi^{(2,1)}_{R}\left(r,\frac{rx}{\mu}\right) = \frac{\mu}{r} \psi^{(1,1)}_{R/(\phi)}\left(r,\frac{rx}{\mu}\right).
\end{equation}
When $\phi = 0$, eqs.\ (\ref{firstABrel}) and (\ref{secABrel}) become
\begin{eqnarray}
\sum_i A_i^{\mu} B_{i \nu} & = & 0 \ , \label{eq:moment1} \\
A_i^{\mu} B_{i \mu} & = & 0 \ . \label{eq:moment2}
\end{eqnarray}

All $SU(2)$ invariant operators may be written as polynomials in
$A_i^{\mu} B_{j \mu}$, $\epsilon_{\mu \nu} A_i^{\mu} A_j^{\nu}$, and
$\epsilon^{\mu \nu} B_{i \mu} B_{j \nu}$.  These polynomials satisfy
various quadratic relations.  Some of these relations can be obtained by contracting
\eqref{eq:moment1} with bifundamental fields (either $A_i^{\mu}$ or $\epsilon^{\mu \nu} B_{i \nu}$ for upper indices, and $B_{i \mu}$ or $\epsilon_{\mu \nu} A_i^{\nu}$ for lower indices).
We can obtain other relations by contracting bifundamental fields with the identity
\begin{equation} \label{eq:epsilon}
\epsilon_{\mu \nu} \epsilon_{\lambda \delta} + \epsilon_{\mu \lambda} \epsilon_{\delta \nu}
+ \epsilon_{\mu \delta} \epsilon_{\nu \lambda} = 0.
\end{equation}
Once we apply \eqref{eq:moment2}, each of these relations will have zero, two, or three
nonzero terms.  We first mod out by the relations that have two terms.
We get two-term relations by contracting \eqref{eq:moment1} with two fields
with different gauge indices, and by contracting \eqref{eq:epsilon} with
four fields with two or three different gauge indices.

Modding out by relations with two terms gives us a seven dimensional toric variety.
The two-term relations are all invariant under a $(\mathbb{C}^*)^7$
action.
We will let the first four charges $q_1,q_2,q_3,q_4$ be the charges
under the $U(1)$ gauge groups.  We choose $q_5$ so that
each contraction of an $A_1$ or $B_1$ with an $A_2$ or $B_2$ and
each contraction of an $A_3$ or $B_3$ with an $A_4$ or $B_4$
has $q_5 = 1$ and all other contractions have $q_5 = 0$.
Likewise, we choose $q_6$ to count the number of $1$-$3$ and $2$-$4$
contractions, and $q_7$ to count $1$-$4$ and $2$-$3$
contractions.
Note that $r = q_5 + q_6 + q_7$.
Also, the following quantitites are even:
\be
q_1+q_2+q_3+q_4 \ , \; \; \; q_1 + q_2 + q_6+q_7 \ , \; \; \;  q_1 + q_3 + q_5 + q_ 7 \ , \; \; \;
q_1 + q_4 + q_5 + q_6  \ .
\ee

We can check that all contractions of a pair of $A$'s and $B$'s
can be expressed in terms of
$\epsilon^{\mu \nu} B_{i \mu} B_{j \nu}$ and
$A_1^{\mu} B_{2 \nu}$.
We obtain an expression for $A_1^{\mu} B_{i \mu}$ by contracting
\eqref{eq:epsilon} with
two $A_1$'s, a $B_2$, and a $B_i$.
We can then obtain $A_j^{\mu} B_{i \mu}$ by contracting \eqref{eq:moment1}
with $B_i$ and $B_k$, where $1,i,j,k$ are all different.
Finally, contracting \eqref{eq:epsilon} with one each of $A_i,B_i,A_j,B_j$ gives
 us $\epsilon_{\mu \nu} A_i^{\mu} A_j^{\nu}$.
Therefore, the variety must be seven dimensional.

For simplicity and to make contact with ``case one'' discussed above, we will assume that
$q_1 \ge q_2 + q_3 + q_4$, $q_2, q_3, q_4 \ge 0$.
By construction, we have the inequalities $q_5, q_6, q_7 \ge 0$.
We claim the following inequalities also hold,
\begin{equation}
\begin{split}
q_5 & \ge q_1 + q_2 - r \ , \\
q_6 & \ge q_1 + q_3 - r \ , \\
q_7 & \ge q_1 + q_4 - r \ .
\end{split}
\end{equation}
Consider the first of these inequalities.  The quantity $r$ is half the total number of $A$'s and $B$'s in the operator.  The quantity $q_1 + q_2$ is the number of $A_1$ and $A_2$ fields that we cannot pair with with $B_1$ and $B_2$ fields respectively.  The $SU(2)$ indices of these excess $A_i$ fields must be contracted in a gauge invariant way. To minimize $q_5$ we can try to contract as many of these excess fields as possible with the $A_3$, $B_3$, $A_4$, and $B_4$ fields.  However, if the excess number of $A_1$ and $A_2$ fields is larger than $r$, then we will still have at least $2(q_1+q_2 - r)$ $A_1$ and $A_2$ fields left over which we will be forced to contract together, yielding the lower bound on $q_5$.

Now we impose the three-term relations that can be derived
from \eqref{eq:moment1} and \eqref{eq:epsilon}.
These relations break the symmetries corresponding to $q_5, q_6, q_7$; only the sum $q_5 + q_6 + q_7$ is preserved.
We will let $\theta(x)$ denote the Heaviside step function, and $\theta_1(x) = x \theta(x)$.
Any monomial that has $q_7$ greater than the minimum
$\theta_1(q_1 + q_4 - r)$ can be reduced to a sum of monomials with
$q_7 = \theta_1(q_1 + q_4 - r)$.

Therefore, the density of operators with a given $q_1, q_2, q_3, q_4, r$ is
\begin{equation}
\begin{split}
\psi_{R/(\phi)}^{(1,1)}(r,q) & = \frac{1}{2} \int dq_5\,dq_6\,\theta(r-q_1) \delta(r-q_5-q_6-\theta_1(q_1+q_4-r)) \\
& \hspace{25mm}  \cdot \theta(q_5) \theta(q_6)
\theta(q_5 - q_1 - q_2 + r) \theta(q_6 - q_1 - q_3 + r) \\
& = \frac{1}{2} \theta(r-q_1) (r - \theta_1(q_1+q_4-r)-\theta_1(q_1+q_2-r)-\theta_1(q_1+q_3-r)).
\end{split}
\end{equation}
The factor of $\frac{1}{2}$ comes from the fact that $q_1 + q_4 + q_5 + q_6$
is always even.
If $q_2 \ge q_3 \ge q_4$, then in piecewise form we have
\begin{equation}
\psi_{R/(\phi)}^{(1,1)}(r,q) =
\left\{
\begin{array}{ll}
\frac{r}{2}, & q_1+ q_2 \le r \\
\frac{2r - q_1 - q_2}{2}, & q_1 + q_3 \le r \le q_1 + q_2 \\
\frac{3r - 2q_1 - q_2 - q_3}{2}, & q_1 + q_4 \le r \le q_1 + q_3 \\
\frac{4r - 3q_1 - q_2 - q_3 - q_4}{2}, & q_1 \le r \le q_1 + q_4 \\
0, & r \le q_1
\end{array}
\right. 
\end{equation}
When we replace $q_i$ with $\frac{rx}{\mu} k_i$ and multiply by
$\mu/r$, we get agreement with the densities computed
by the matrix model.

We may also compare the $\delta y_{a,I}$ values with numbers of operators in the chiral ring.
A conjecture in \cite{Gulotta:2011si} states that for a bifundamental field $X_{ab}$ that transforms in the anti fundamental of the $a$'th $U(N)$ gauge group and the fundamental of the $b$'th group, 
\be
\psi_{R / (X_{ab})}^{(1,1)} (r, rx/\mu) = \frac{r}{\mu} \rho(x) [y_b(x) - y_a(x) + R(X_{ab}) ]
\ee
where $R(X_{ab})$ is the R-charge of $X_{ab}$.   In this case, $R(X_{ab}) = 1/2$ because of the ${\mathcal N}=3$ SUSY.

First, we count the operators that are not divisible by
$\epsilon_{\mu \nu} A_1^{\mu} A_2^{\nu}$.
Dividing by $\epsilon_{\mu \nu} A_1^{\mu} A_2^{\nu}$ decreases $q_5$ by one.
Since all operators have $q_5 \ge 0$, any operator with $q_5 = 0$ is not
divisible by $\epsilon_{\mu \nu} A_1^{\mu} A_2^{\nu}$.  We can check that
all other operators are divisible by $\epsilon_{\mu \nu} A_1^{\mu} A_2^{\nu}$.

Since $q_5 \ge q_1 + q_2 - r$, operators with $q_5 = 0$
exist only if $r \ge q_1 + q_2$.  Therefore,
\begin{eqnarray}
\psi^{(1,1)}_{R/(\phi,\epsilon_{\mu \nu} A_1^{\mu} A_2^{\nu})}(r,q)
& = & \frac{1}{2} \theta(r - q_1 - q_2) \ ,  \\
\psi^{(1,1)}_{R/(\epsilon_{\mu \nu} A_1^{\mu} A_2^{\nu})}(r,q)
& = & \frac{1}{2} \theta_1(r - q_1 - q_2) \ . 
\end{eqnarray}
The factor of $\frac{1}{2}$ comes from the fact that
$q_1 + q_2 + q_6 + q_7 = q_1 + q_2 + r$ must be even.
Therefore, operator counting predicts that
\begin{equation}
\rho(x) (1 - y_1 - y_2 + y_{0,1} + y_{0,2}) = \frac{1}{2} \theta_1(\mu - (k_1 + k_2)x).
\end{equation}
We can check that this agrees with our matrix model calculations.
Similarly,
\begin{eqnarray}
\rho(x) (1 - y_1 - y_3 + y_{0,1} + y_{0,2}) & = & \frac{1}{2} \theta_1(\mu - (k_1 + k_3)x) \ , \\
\rho(x) (1 - y_1 - y_4 + y_{0,1} + y_{0,2}) & = & \frac{1}{2} \theta_1(\mu - (k_1 + k_4)x).
\end{eqnarray}

Finally, we count the operators that are not divisible by $A_1^{\mu} B_{2 \mu}$.
Dividing by $A_1^{\mu} B_{2 \mu}$ decreases $q_1$, $q_5$ and $r$ by one
and increases $q_2$ by one.  Since $q_5$ and $q_5 - q_1 - q_2 + r$ must both
be nonnegative, operators with $q_5 = 0$ or $q_5 - q_1 -q_2 + r$ are not
divisible by $A_1^{\mu} B_{2 \mu}$.

There are two cases to consider.  If $q_5 = 0$, then for $r \geq q_1 + q_2$, the value of 
$\psi^{(1,1)}_{R / (\phi,A_1^{\mu} B_{2\mu})}(r,q)$ must be 1/2 as we had before.  
The second case to consider is $q_5 = q_1 + q_2 - r  > 0$.
This constraint implies that $q_1 + q_2 + q_6 + q_7 = (q_1 + q_2 - q_5) + (q_5 + q_6 + q_7) = 2r$
is even and we do not need an additional factor of 1/2.  
All operators must satisfy the constraint $r > q_1$.  Thus in the range
$q_1 < r < q_1 + q_2$, the value of $\psi^{(1,1)}_{R / (\phi,A_1^{\mu} B_{2\mu})}(r,q)$ must be one.
In summary, we find
\begin{eqnarray}
\psi^{(1,1)}_{R/(A_1^{\mu} B_{2 \mu})}(r,q)
& = & \theta_1(r -q_1) - \frac{1}{2} \theta_1(r - q_1 - q_2) \ .
\end{eqnarray}
Operator counting predicts that
\begin{equation}
\rho(x) (1 - y_1 + y_2) = \theta_1(\mu - k_1 x) - \frac{1}{2} \theta_1(\mu - (k_1 + k_2)x) \ ,
\end{equation}
and this agrees with our matrix model results.

\section{Discussion}

In this paper, we saw that there is an ADE classification of
 ${\mathcal N}=3$ SUSY CS theories in 2+1 dimensions with bifundamental matter fields whose partition function $Z_{S^3}$ has a nice large $N$ limit.  By nice limit, we mean that when we use the matrix model of \cite{Kapustin:2009kz} to calculate $Z_{S^3}$, there is cancellation in the long range forces between the eigenvalues.  The classification raises a number of questions which we were not able to answer in this paper.  Leading among them is the significance of the $N_f = 2 N_c$ condition (\ref{ADEcond}).  For a 3+1 dimensional gauge theory, this condition implies the vanishing of the NSVZ beta function and the existence of a conformal fixed point, but in the present context the meaning is obscure.  It is tempting to speculate that $N_f = 2N_c$ is necessary in 2+1 dimensions for the existence of the fixed point.  However, it could be that the condition is just a technical limitation of the method introduced in \cite{Herzog:2010hf} for calculating the large $N$ limit of the matrix model.

\begin{figure}
\begin{center}
\epsfig{file=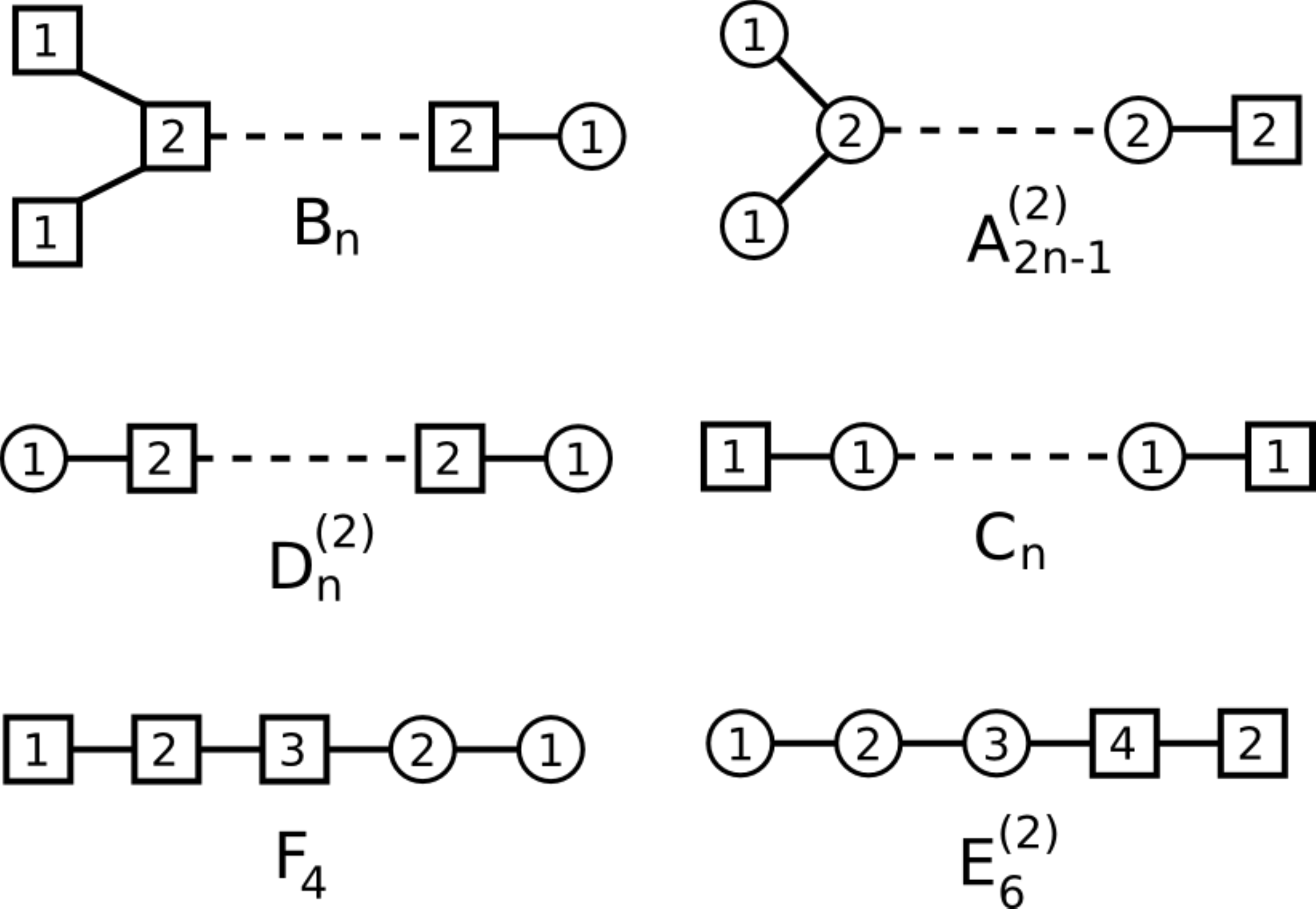,width=4in,angle=0,trim=0 0 0 0}
\end{center}
\caption{\label{fig:sosp}We conjecture that these are the six families of
quivers involving both unitary and orthogonal/symplectic gauge groups that
have $\mathcal{N}=3$ supersymmetry.  The circular nodes correspond to
unitary groups, and the square nodes correspond to orthogonal or symplectic
groups.  These quivers can be interpreted as non-simply-laced
extended Dynkin diagrams.  An edge
connecting a unitary gauge group to an orthogonal/symplectic gauge group
corresponds to a double bond in the Dynkin diagram, with the arrow pointing
toward the unitary gauge group.  We are not sure of the precise rules for
deciding which gauge groups should be orthogonal and which should be symplectic.
There does not appear to be a way to realize the extended Dynkin diagrams 
$G_2$, $I_1$, $A_{2n-1}^{(2)}$, or $D_4^{(3)}$ as quivers,
since we do not have a way of interpreting triple or $\infty$ bonds, or a pair of double bonds pointed in the same direction.}
\end{figure}

Another question has to do with the restriction to $U(N)$ gauge groups.  It would be interesting to study quivers of $SO(N)$ and $Sp(N)$ groups in the large $N$ limit.  We conjecture that ``good'' theories can all be described by extended Dynkin diagrams.  We believe that the ``good'' quivers containing just $SO(N)$ and $Sp(N)$ gauge groups are simply laced extended Dynkin diagrams,
while the ``good''
quivers that also contain $U(N)$ are the ones shown in Figure \ref{fig:sosp}.
It would be interesting to study this correspondence further.
We note that an $SO$-$Sp$ version of the simply laced $A_1$ quiver was
studied in \cite{Aharony:2008gk}.
 
We would like to end this paper by describing some progress we made in writing $F$ in a way that is manifestly invariant under Seiberg duality.  
In the introduction, we reviewed the fact that $\Vol(Y) \sim 1/F^2$ where $Y$ is the tri-Sasaki Einstein space in the dual eleven dimensional supergravity description of the CS theory.
In the $A_n$ and $D_4$ cases, we have found that $\Vol(Y)$ can be written as a rational function of the CS levels $k_a$.  The numerator of this function takes the form
\be
\sum_{\alpha_1, \ldots, \alpha_n \in R} \det(\alpha_1, \ldots, \alpha_n)^2  \prod_{a=1}^n 
|\alpha_a \cdot p|
\ee
where $R$ is the set of roots of the $A_n$, $D_n$, or $E_n$ Lie algebra.  Recall that the CS levels can be reconstructed from the simple roots via $k_a = \alpha_a \cdot p$.

\section*{Acknowledgments} We would like to thank D.~Berenstein, I.~Klebanov, J.~McGreevy, T.~Nishioka, S.~Pufu, M.~Ro\v{c}ek, and E.~Silverstein for discussion. This work was supported in part by the US NSF under Grants No. PHY-0844827, PHY-0756966, and
PHY-0551164.  CH thanks the KITP for hospitality and the Sloan Foundation for partial support.

\appendix

\section{A Simple Class of Solutions}
\label{app:simple}

In this section, we find a simple class of solutions for any ADE quiver.
Let us start by looking at the $E_6$ quiver.  

At the risk of unacceptably proliferating the number of indices, we label the CS levels of the external nodes $k_{1a}$, of the internal nodes $k_{2a}$, and of the central node $k_3$ with the constraint $\sum_a (k_{1a} + 2 k_{2a} ) = -3k_3$.  The imaginary parts of the eigenvalues are 
correspondingly labeled $y_{1a}$, $y_{2a,I}$ and $y_{3,J}$.  
For small enough $x$, the equations of motion yield
\be
\rho(x) = \frac{\mu}{6} \ ; \; \; \;
y_{1a} - y_{2a,I} = \frac{3 k_{1a} x}{\mu} \ ; \; \; \;
y_{2a,I} - y_{3,J} = \frac{(k_{1a} + 2 k_{2a})x}{\mu} \ .
\ee
By small enough, we mean that no difference between the $y$'s has saturated to $\pm 1/2$.  More precisely, we require $| x | \leq \mu / 6 \abs{k_{1a}}$ and $|x | \leq \mu / 2 \abs{k_{1a} + 2 k_{2a}}$ for all $a$.

As in the $D_4$ case, for a certain class of solutions the eigenvalue distribution ends inside this central region.  Let us try to extremize the free energy as a function of the endpoint $x_*$.  The on shell free energy as a function of $x_*$ is
\be
F(x_*) = N^{3/2} \left[ \frac{3 \pi}{x_*} + \frac{ \pi x_*^3}{9}\sum_a \left(3 k_{1a}^2 + (k_{1a}+2k_{2a})^2 \right) \right]  \ .
\ee
which leads to the condition that 
\be
x_* = \sqrt{3} \left[\sum_a \left(3 k_{1a}^2 + (k_{1a}+2k_{2a})^2 \right)\right]^{-1/4} \ .
\ee
In order for $x_*$ to lie in the central region, we must take $k_{1a} = k_{2b} = \pm k$ or $2 k_{1a} = - k_{2b} = \pm 2k$, leading to the final result
\be
F = \pi N^{3/2} 4 \sqrt{2 \abs{k}} \ ,
\ee
and the result for the volume
\be
\operatorname{Vol}(Y) = \frac{\pi^4}{432 |k| } \ .
\ee
The tri-Sasaki Einstein manifold in question is a level surface of the hyperk\"ahler quotient
${\mathbb H}^{24} /// SU(3) \times SU(2)^3 \times U(1)^5$.

The solution above has some characteristic features that we can try to generalize for arbitrary ADE quivers.  The $y_{1a} - y_{2a,I}$ and $y_{2a,I} - y_{3,J}$ all saturate to $\pm 1/2$ at $x=x_*$.  Also, the absolute values of the $\delta y$ are equal,
$|y_{1a} - y_{2a,I}| = |y_{2b,I} - y_{3,J}|$.
For the arbitrary ADE case, let the CS level of the $U(n_a N)$ gauge group be $k_a = \mbox{sgn}(a) k n_a$, where we choose $\mbox{sgn}(a) = \pm 1$ so that neighboring gauge groups have opposite sign.  
 This choice of CS levels is consistent with the condition $\sum_a n_a k_a = 0$.  The equations of motion following from (\ref{finalF}) are satisfied if we set all the $|\delta y|$ equal to each other and the eigenvalue density to a constant:
\be
\delta y_{ab,IJ} = \mbox{sgn}(a) \frac{x k}{2 \rho} \ , \; \; \; \rho = \frac{4 \mu}{\sum_a n_a^2} \ .
\ee
With a little more work, we can find the chemical potential and the volume of the associated tri-Sasaki Einstein manifold
\be
\operatorname{Vol}(Y) = \frac{\pi^4}{24 \mu^2} = \frac{\pi^4}{3|k|} \left( \frac{2}{ \sum_a n_a^2} \right)^2 \ .
\ee

\end{document}